\documentclass{article}
\usepackage{spconf,amsmath,graphicx,amssymb}
\graphicspath{{./images_feb7/jan_31_as20_asw5_nr1000/}}
\usepackage{enumitem}
\setlist{nosep, leftmargin=14pt}

\usepackage{mwe} 

\def\pr{\text{pr}}

\def\L{\mathbb{L}}
\def\E{\mathbb{E}}
\def\R{\mathbb{R}}
\def\H{\mathcal{H}}
\def\f{\mathbf{f}}
\def\g{\mathbf{g}}
\def\n{\mathbf{n}}
\def\tht{\pmb{\theta}}
\def\th{\mathrm{th}}
\def\b{\boldsymbol{b}}
\def\thetab{\pmb{\theta}}
\def\etal{\textit{et al.}}
\title{Ideal-observer computation with anthropomorphic phantoms using Markov Chain Monte Carlo}
\name{Md Ashequr Rahman$^{1,2}$ \quad Zitong Yu$^{1, 2}$ \quad Abhinav~K.~Jha$^{1,2}$}

\address{$^1$Department of Biomedical Engineering, Washington University in St. Louis, St. Louis, MO, USA\\
$^2$Mallinckrodt Institute of Radiology, Washington University in St. Louis, St. Louis, MO, USA}
\begin{document}
%
\twocolumn[
  \begin{@twocolumnfalse}
  \section*{IEEE Copyright Notice:}
   \textcopyright 2022 IEEE.  Personal use of this material is permitted.  Permission from IEEE must be obtained for all other uses, in any current or future media, including reprinting/republishing this material for advertising or promotional purposes, creating new collective works, for resale or redistribution to servers or lists, or reuse of any copyrighted component of this work in other works.
   
   \vspace{100pt}
   Accepted to be published in: 2022 IEEE International Symposium on Biomedical Imaging (IEEE ISBI 2022), March 28-31, 2022
  \end{@twocolumnfalse}
  ]
\clearpage

\maketitle
\begin{abstract}
In medical imaging, it is widely recognized that image quality should  be objectively evaluated based on performance in clinical tasks.
To evaluate performance in signal-detection tasks, the ideal observer (IO) is optimal but also challenging to compute in clinically realistic settings. 
Markov Chain Monte Carlo (MCMC)-based strategies have demonstrated the ability to compute the IO using pre-computed projections of an anatomical database. 
To evaluate image quality in clinically realistic scenarios, the observer performance should be measured for a realistic patient distribution. This implies that the anatomical database should also be derived from a realistic population. In this manuscript, we propose to advance the MCMC-based approach to achieve these goals. 
We then use the proposed approach to study the effect of anatomical database size on IO computation for the task of detecting perfusion defects in simulated myocardial perfusion SPECT images.
Our preliminary results provide evidence that the size of the anatomical database affects the computation of IO.
\end{abstract}
\begin{keywords}
Image-quality evaluation, SPECT, Markov chain Monte Carlo.  
\end{keywords}
\section{Introduction}
\label{sec:intro}
There is wide recognition in the medical imaging community that image quality should be evaluated based on performance in clinical tasks, such as those of detection and estimation \cite{barrett2015task}.
Of these, detection tasks are performed by an observer, who makes a decision on whether the signal to be detected is present or absent.
The term observer typically conjures the images of a trained radiologist, but can also be mathematical observers, also referred to as model observers \cite{barrett1993model}. 
Of the various observers, the one that utilizes all the statistical information available regarding the task to maximize task performance is referred to as the ideal observer (IO) \cite{barrett1998objective}.
This observer provides the best possible performance on the detection task, as quantified using the area under the receiver operating characteristic (ROC) curve (AUC).
The use of IO is recommended for optimizing system instrumentation and acquisition protocols to ensure that the measured data has the maximum possible information for the detection task \cite{jha2021objective, barrett2015task}. 
However, the IO requires a complete knowledge of the distribution of the image data. 
Such a distribution is very high dimensional, and, thus, very difficult, if not, impossible to obtain. 
To address this challenge, various approaches have been proposed \cite{clarkson2010fisher,jha2013ideal} but computing IO performance in clinically realistic settings remains a challenge.  
Thus, there is an important need for strategies to compute the IO in clinically realistic scenarios.

A seminal contribution towards IO computation was a Markov Chain Monte Carlo (MCMC)-based technique proposed by Kupinski \etal~\cite{kupinski2003ideal}.
The technique demonstrated the ability to compute IO for parametric object models, in particular, a lumpy background model. 
This technique was later extended for a simple parametric description of the human cardiac region~\cite{he2008toward}.
To improve the efficiency of the IO computation, projection data of an anatomical database were precomputed.
To increase the realism of this approach, Ghaly \etal~\cite{ghaly2015optimization} proposed a strategy that used anthropomorphic extended cardiac and torso (XCAT) phantoms \cite{segars20104d} that simulated the anatomical variability in patient populations. 
The above studies focused on IO computation on a patient population with anatomical parameters sampled from a uniform distribution.
However, in clinical settings, the distribution of anatomical parameters is unlikely to be uniformly distributed. For example, patient heights typically follow normal distributions \cite{komlos1990estimating}.
For accurate computation of the IO, and in general, for rigorous image-quality evaluation studies, the anatomical variability needs to be sampled from  realistic patient populations. 
Thus, the use of uniform distributions to sample anatomical parameters may not reflect IO performance in clinically realistic settings. 

Another important consideration in using MCMC-based approaches to IO computation is the size of the anatomical database. He \etal.~\cite{he2008toward} indicated that this size may affect the calculation of IO. Further, they suggested that the number of anatomies required to accurately compute the IO may be a function of system resolution. Thus, the effect of the size of anatomical database on the computation of IO also needs to be investigated.


In this paper, we advance the previously developed MCMC-based methods to achieve the goal of performing IO computation with clinically realistic populations. 
In this proof-of-concept study, we show the ability of the proposed method to sample from an anatomical database where the distribution of height and heart sizes are described by clinically realistic normal distributions. 
We then validate the use of this method to compute the IO performance. 
We further use the proposed method to investigate the effect of size of the anatomical database on the accuracy of the IO computation. 
We implement and evaluate the method in the context of computing the IO for the task of detecting cardiac defects from myocardial perfusion SPECT images. 

\section{Theory}
\label{sec:method}
Consider a myocardial perfusion SPECT system imaging a radiotracer distribution, denoted by the infinite-dimensional vector $\f$, that, we assume, lies in the Hilbert space $\L_2(\R^3)$. 
Denote the SPECT imaging system by the Hilbert space operator $\H$. 
Next, denote the projection data obtained by the SPECT system by the $M$-dimensional vector $\g$, that, we assume, lies in the Hilbert space $\E^M$. Thus the SPECT system can be described by the mapping $\H:\L_2(\R^3) \rightarrow \E^M$, and the imaging-system equation is given by
\begin{equation}
    \g = \H \f + \n,
\end{equation}
where $\n$ denotes the $M$-dimensional noise vector. 
We write the object $\f$ as a combination of the signal of interest, denoted by $\f_s$ and the rest of the object (referred to as the background object), denoted by $\f_b$. Thus
\begin{equation}
    \f = \f_s + \f_b.
\end{equation}
In a clinical setting, $\f_s$ could denote the region with abnormal uptake, such as a lesion, while the anatomical and physiological variability in the rest of the patient would be denoted by $\f_b$.
Finally, let the noise-free image corresponding to the background object be denoted by $\b$, i.e.~$\b = \H \f_b$.

In a detection task performed on the projection data $\g$, the goal is to determine, from the projection data, whether the underlying signal of interest is present or absent.
Let $\pr(\g|H_1)$ and $\pr(\g|H_0)$ denote the probability distribution function of the image given that the signal is present and absent, respectively. 
To compute the IO, the following test statistic, referred to as the likelihood ratio, and denoted by $\Lambda(\g)$, is computed:
\begin{equation}
    \Lambda(\g) = \frac{\pr(\g|H_1)}{\pr(\g|H_0)}.
    \label{eq:lr}
\end{equation}
From this equation, we observe that computing the IO requires complete knowledge of the distribution of $\g$ under both the signal present and signal-absent hypothesis. 

When the to-be-detected signal $\f_s$ and background $\f_b$ are both known exactly, this test statistic can be computed using the knowledge of the noise statistics. 
We denote this test statistic by $\Lambda_{BKE}(\g)$.
However, in a clinically realistic setting, where both the signal and the background vary, the probabilities in Eq.~\eqref{eq:lr} can be difficult, if not impossible to define. 
A simplification is obtained when we consider the detection task where the properties of the  signal are known, while the rest of the patient properties are variable. 
For this detection task, referred to as the signal-known-exactly/background known statistically (SKE/BKS) task, the expression for the IO is given by \cite{kupinski2003ideal}:
\begin{equation}
    \Lambda(\g) = \int d \b \Lambda_{BKE}(\g | \b ) \pr(\b| \g, H_0).
\end{equation}
If we could sample backgrounds from the posterior distribution $\pr(\b| \g, H_0)$, then this integral can be computed through a Monte Carlo integration procedure:
\begin{equation}
    \Lambda(\g) = \frac{1}{J} \sum_{j=1}^J \Lambda_{BKE}(\g|\b^j),
\end{equation}
where $\b^j$ denotes the $j^{\th}$ realization of the background.
However, $\b$ is a very high-dimensional vector (~$128^3$ for SPECT). Sampling from this distribution is challenging. An alternative is to consider a parametric representation of the background \cite{kupinski2003ideal, he2008toward}.
However, human anatomies are challenging to represent through parametric models. 
To address this issue, Ghaly \etal ~\cite{ghaly2015optimization} proposed the use of XCAT phantoms. 
However, to the best of our knowledge, their anatomical database size was limited to $N_d$ = 54.
However, larger-sized databases may be needed to conduct rigorous image-quality evaluation studies.
Moreover, in their approach, the anatomical parameters were sampled from a uniform distribution, which, as mentioned above, has limitations in modeling realistic populations. 

In this study, we advance the above MCMC-based approach to sample from a more realistic patient population distribution. 
Similar to He \etal, we parameterize the anatomy of each sample of the patient population using two parameters, namely, the radius of the left ventricle of the heart, and the body size. These are denoted by $\theta_h$ and $\theta_b$, respectively. 
The activity uptake in the different organs, including the heart, liver and lungs, is parameterized by a q-dimensional vector $\thetab_{act}$. 
Denote $\thetab = \{\thetab_{act}, \theta_h, \theta_b$\} and the $i^{\th}$ element of $\thetab$ in iteration $j$ by $[\thetab]_i^j$ . 
Mathematically, at each iteration, we sample from a one-dimensional proposal distribution given by:
\begin{align}
    \tilde{[\thetab]}_i \sim \mathcal{N}([\thetab]_i^j, \sigma_i),
\end{align}
where $\mathcal{N}$ denotes a normal distribution with mean $[\thetab]_i^j$ and standard deviation $\sigma_i$. 
Further, $\tilde{[\thetab]_i}$ denotes the newly proposed sample of the $i^{\th}$ parameter. 
Note that for the $j^{\th}$ iteration, 
\begin{align}
    &i \sim \mathcal{U}(1,q+2),\\
    &\tilde{[\thetab]}_m = [\thetab]_m^j \quad ,\forall m\neq i,
\end{align}
where, $\mathcal{U}(1,q+2)$ denotes sampling from a discrete uniform distribution with range between 1 to $q+2$ with a step size of 1.
As the anatomical database contains only discrete LV radius and body size, the anatomy proposed at each iteration should be chosen such that the LV radius and body size of that anatomy are closest to the proposed LV radius and body size.  
This strategy provides a way to guide the sampling process through a large-sized phantom population and allows for the prior information about the LV radius and body size to be accounted for in the calculation of acceptance ratio.

\section{Implementation and evaluation of the proposed approach}
\label{sec:experiment}
\subsection{Anthropomorphic phantom model and projection-data generation}
We used the XCAT phantom \cite{segars20104d} for generating patient populations. 
For realistic variability in phantom, the body size, LV length and radius, and the body height were sampled from independent Gaussian distribution with mean and variance based on Emory PET dataset and listed in Ghaly \etal~\cite{ghaly2014design}.
We only considered one transaxial slice of each anatomy that contained the centroid of the defect in defect-present case.
The phantom was generated over a $256 \times 256$ grid with pixel size of $2$ mm. The defect-present case was generated by introducing a cold defect with $25\%$ severity and $10\%$ extent in the anterior wall. 
A total of $10000$ pairs (defect present/absent) of anatomies were generated, $5000$ pairs each for both male and female patients. 

Projection data for the phantom population was generated with a simulated 2-D SPECT system that modeled the collimator-detector response and noise in SPECT systems. 
The projection data was acquired in $64$ projection bins and at $3$ projection angles. 
The pixel size of each projection bin was $8~\text{mm}$. 
As in \cite{ghaly2015optimization} and \cite{he2008toward}, we pre-calculated the organ-specific projection for the $10,000$ anatomies for defect-absent case. The defect-only projection was also computed to be used in SKE/BKS task and to generate the defect-present projections to be used in observer study. 
The organs considered were the heart, lung, liver and the background.
This strategy of obtaining organ-specific projections substantially reduced the computational requirements of the MCMC method. 

\subsection{Generating the population for observer study}
We used the generated projection data from the 10,000 anatomies and defect-only projection for each anatomy to generate 2000 pairs of projection data, $\boldsymbol{g}$ using as similar strategy as in \cite{he2008toward}. In brief, for organ $i$ and anatomy index $k$, denote the noise-free projection data as $\b_{i,k}$. Denote the LV radius and body size associated with anatomy index $k$ as $\theta_{h,k}$ and $\theta_{b,k}$, respectively. Then the noise-free phantom projection data is given by
\begin{align}
    \b(\{\thetab_{act}, \theta_{h,k}, \theta_{b,k}\}) = \sum_{i=1}^{4}[\thetab_{act}]_i\b_{i,k}. \label{eq:b}
\end{align}
We refer to this dataset as the test set for computing IO performance.

We scaled the background projection data such that, on average, the total number of counts in projection data was 5000. 
To get the final projection data, $\g$, we added Poisson noise to this projection data. For each anatomy, we sampled the organ activity from distribution mentioned in \cite{ghaly2014design}.

\subsection{The MCMC simulation}
We initialized the activity parameters by randomly sampling from the distribution given in  \cite{ghaly2014design}. 
The initial anatomy parameter was sampled uniformly from the anatomical database. 
Then at each iteration of MCMC, we randomly picked a single parameter to modify. 
For proposing a new activity parameter $[\thetab_{act}]_i$, we used a Gaussian proposal density with standard deviation set to one-tenth of the standard deviation of that particular prior organ activity distribution. 
However, to sample from the anatomy parameter $k$, we made use of the distribution of LV radius and body size. 
Based on the previous iterations' anatomy $k^j$, we obtained the corresponding LV radius and body size. 
For proposing a new LV radius, we set the proposal density as a Gaussian distribution. The standard deviation of this distribution was one-fifth the standard deviation of the clinically observed LV-radius distribution. 
Similar strategy was used for proposing new body size. Using these proposed parameters, we selected the anatomy for which the LV radius and body size was the closest match. As the proposal density is symmetric, we accepted the newly proposed parameter set $\tilde{\tht}$ over the previous parameter $\tht^{j}$ with probability
\begin{align}
    \alpha(\tilde{\tht}, \tht^{j}) = \min\left(1, \frac{\pr(\g|\tilde{\tht},H_0)\pr(\tilde{\tht})}{\pr(\g|\tht^j,H_0)\pr(\tht^j)}\right).
\end{align}
Note that, prior probabilities can be modeled again by a Gaussian distribution from the patient data.

Challenges in implementation of the MCMC technique include cases where some projection bins in $\tilde{\b}=\b(\tilde{\thetab})$ and $\b^j=\b({\thetab^j})$ are null. This can be addressed by discarding the contribution of such pixels while calculating the acceptance probability. 
We also discard the initial 400,000 burn-in iterations where non-stationarity of $\Lambda_{BKE}$ estimates are observed to get a reliable IO estimate.    

\section{Results}
\label{sec:results}
\begin{figure}[htb]

\begin{minipage}[b]{.48\linewidth}
\centering
\centerline{\includegraphics[width = 4.2 cm]{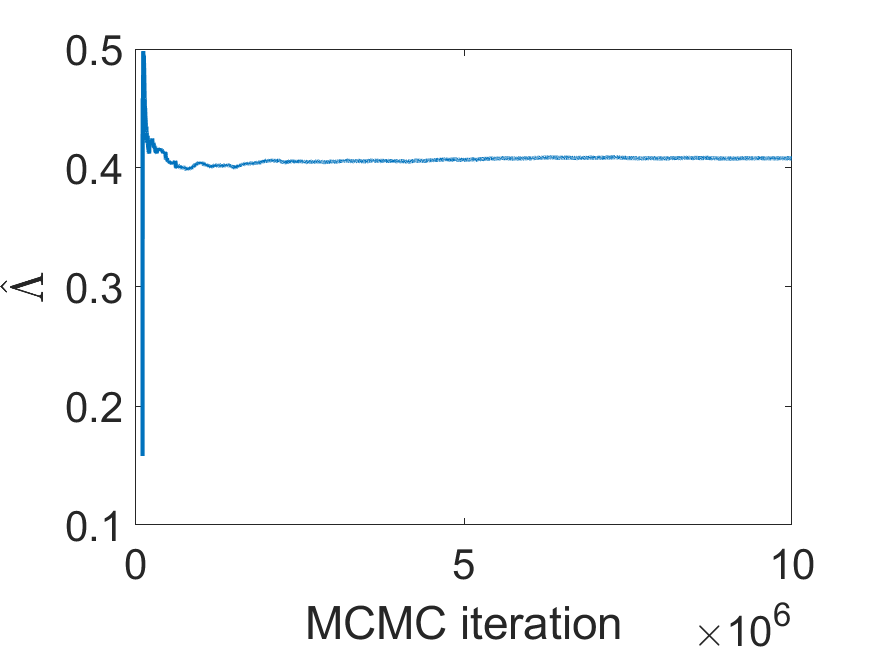}}
\centerline{(a)}\medskip
\end{minipage}
\hfill
\begin{minipage}[b]{0.48\linewidth}
\centering
\centerline{\includegraphics[width = 4.2 cm]{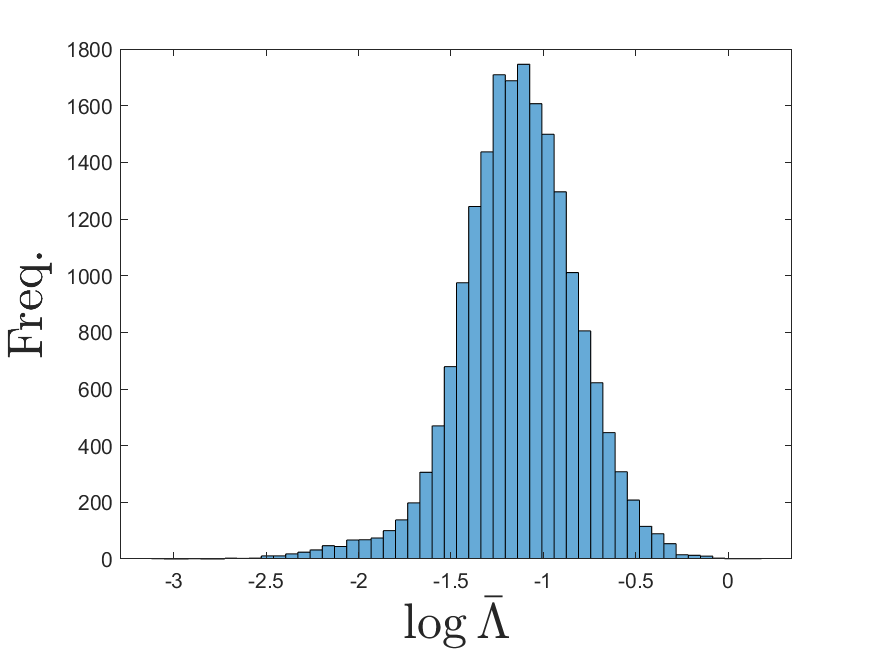}}
\centerline{(b)}\medskip
\end{minipage}

\caption{(a) The estimate of IO test statistics as a function of total number iterations used for MCMC simulation. (b) The distribution of block estimate of log of IO test statistic.}
\label{fig:fig_1}
\end{figure}

In Fig.~\ref{fig:fig_1}(a), we show the convergence of IO estimate when the size of anatomical database, $N_d$ was $10000$. We observed that the estimate converges after around 4 million iterations. We also divide the entire iterations into consecutive blocks and show the estimate of IO test statistics calculated from each block, as in \cite{he2008toward}. The block size was set to 500 iterations. We observed that the distribution follows a log-normal distribution  and the log of this estimate follows a normal distribution (Fig.~\ref{fig:fig_1}(b)). 

\begin{figure}[htb]

\begin{minipage}[b]{.48\linewidth}
  \centering
  \centerline{\includegraphics[width=4.2cm]{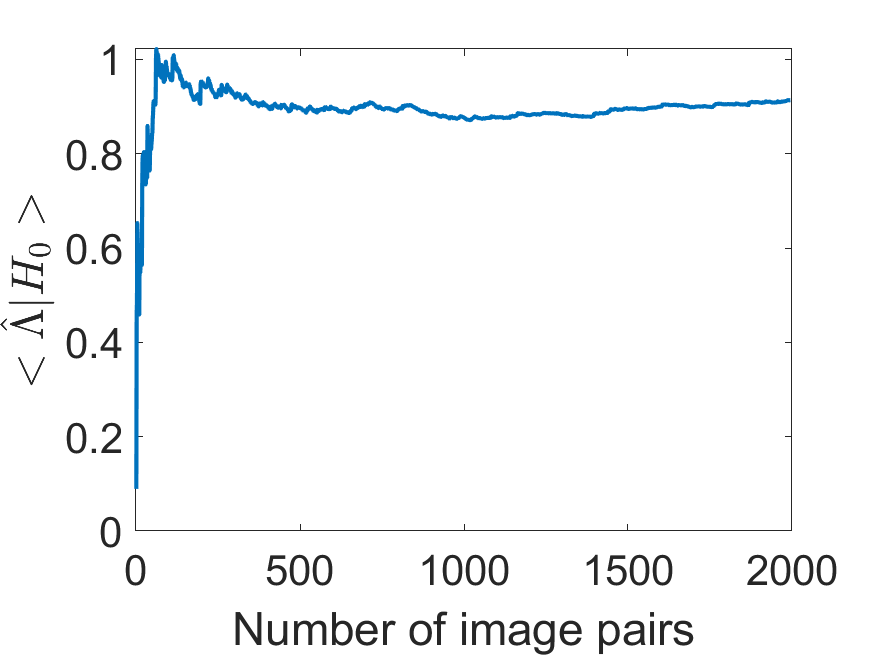}}
  \centerline{(a)}\medskip
\end{minipage}
\hfill
\begin{minipage}[b]{0.48\linewidth}
  \centering
  \centerline{\includegraphics[width=4.2cm]{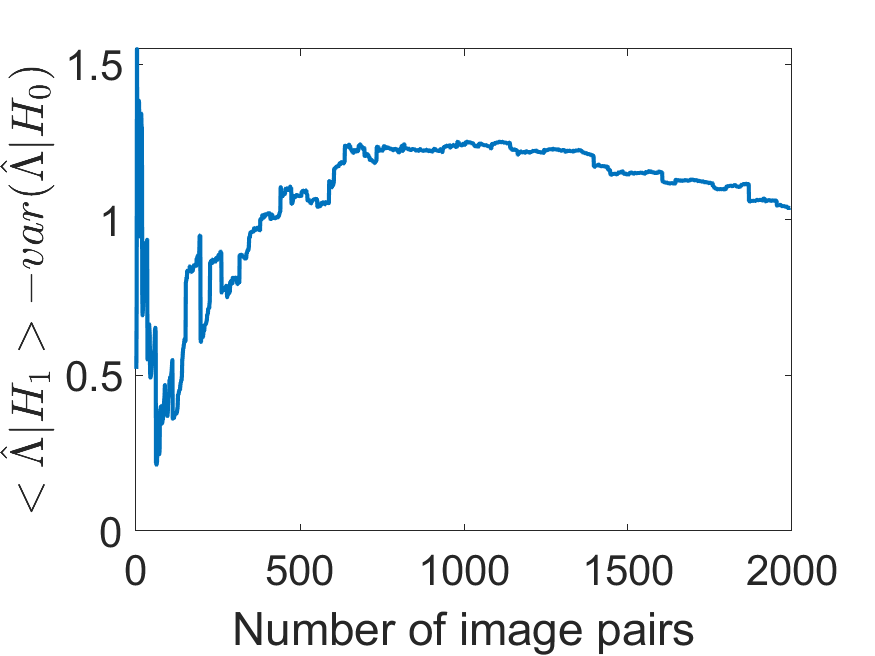}}
  \centerline{(b)}\medskip
\end{minipage}
\caption{(a) $\langle\hat{\Lambda}|H_0\rangle$ and (b) $\langle\hat{\Lambda}|H_1\rangle-var(\hat{\Lambda}|H_0)$ as a function of number of image pairs used for observer study.}
\label{fig:ensemble_plots}
\end{figure}

We also validate the proposed MCMC method by comparing empirical and theoretical estimate of $\langle\Lambda|H_0\rangle$ and $\langle\Lambda|H_1\rangle - Var(\Lambda|H_0)$. Theoretically, both these terms should be equal to one \cite{he2008toward}. From Fig.~\ref{fig:ensemble_plots}, we observed that these quantities reach $0.913$ and $1.03$, respectively  for $N_d = 10000$. 
The closeness of these summary statistics to 1 provides evidence in the direction of validating this method.
The minor deviation could be due to the unique nature of distribution of anatomical phantom population and needs further investigation. 
However, overall, the results in Figs.~\ref{fig:fig_1} and \ref{fig:ensemble_plots} provide evidence in support of validating the MCMC technique. 

\begin{figure}[htb]
\centering
\includegraphics[width = 2.5 in]{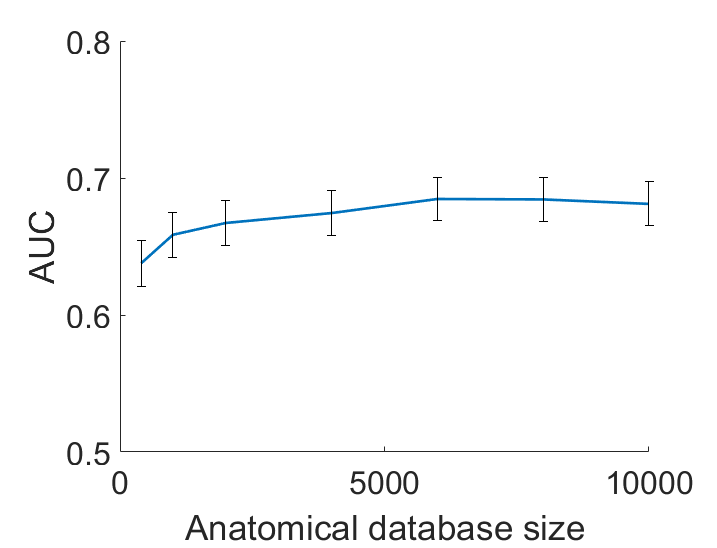}
\caption{AUC values as a function of size of anatomical database. The error bar indicates $95\%$ confidence interval.}
\label{fig:auc_size}
\end{figure}

We next investigated the effect of the size of anatomical distributions on IO computation using the MCMC technique. For this purpose, we varied the size of anatomical database and for each setting, computed the IO performance. The AUC value as a function of anatomical database size is shown in Fig.~\ref{fig:auc_size}. We observed that increasing the size of anatomical database resulted in an increase in AUC value. Thus, a large-sized population of the anatomical database is required to get reliable estimate of the IO performance through the MCMC approach for anthropomorphic phantoms. 

\section{Discussions and Conclusion}
\label{sec:discussion}

In this paper, an MCMC-based method was proposed in the context of computing IO performance in clinically realistic settings.
We observe that the proposed method operates with the anthropomorphic XCAT phantoms, and is able to sample the anatomical descriptors of this phantom from a clinically realistic normal distribution. 
Our investigation of the effect of size of anatomical database on IO computation with this method showed that large sample size of the anatomical patient database are required to accurately compute the IO performance. These findings thus suggest the use of approaches that can generate such large databases. 

The proposed approach has multiple applications in addition to computing the IO. This includes generating phantom populations for virtual clinical trials and other image-quality evaluation studies \cite{barrett2008adaptive, ghanbari2017optimization, yu2021physics, liu2021bayesian, yu2020ai}. Another application is in simulation-guided deep learning approaches that provide the advantage of using patient populations with known ground truth, which can then be used for training \cite{leung2020physics}. The proposed method may be used to generate such a patient population.

In conclusion, we advanced a MCMC-based strategy with the eventual goal of sampling from a clinically realistic patient population. 
We then demonstrated the application of this strategy to sample from an anatomical database with height and heart sizes described by clinically realistic normal distributions. 
The strategy was validated in the context of evaluation IO performance for defect-detection tasks in myocardial perfusion SPECT. 
Our analysis provides evidence of the efficacy of having a large-sized anatomical database to reliably compute IO performance.


\section{Compliance with ethical standards}
\label{sec:ethics}
This is a numerical simulation study for which no ethical approval was required.

\section{Acknowledgments}
\label{sec:acknowledgments}
This work was supported by the National Institute of Biomedical Imaging and Bioengineering of the National Institute of Health under grants R21-EB024647, 
R01 EB031051, and R56 EB028287. 
\bibliographystyle{IEEEbib}
\bibliography{strings,refs}

\end{document}